\documentclass[aps, prb, reprint, showpacs,amsmath,amssymb,superscriptaddress,floatfix]{revtex4-2} 
\usepackage{braket}
\usepackage{cancel}
\usepackage{titlesec}
\usepackage[utf8]{inputenc}
\usepackage[dvipdfmx]{graphicx}
\usepackage{here}
\usepackage[dvipdfmx]{color}
\usepackage{dcolumn}
\usepackage{enumerate}
\usepackage{csquotes}

\begin{document}
\title{Dual role of longitudinal optical phonons for generation of coherent oscillations in gallium arsenide under optical pumping}

\author{Itsuki Takagi}
\affiliation{Laboratory for Materials and Structures, Institute of Innovative Research, Tokyo Institute of Technology, 4259 Nagatsuta, Yokohama 226-8501, Japan}
\affiliation{Department of Materials Science and Engineering, Tokyo Institute of Technology, 4259 Nagatsuta, Yokohama 226-8501, Japan}

\author{Yuma Konno}
\affiliation{Laboratory for Materials and Structures, Institute of Innovative Research, Tokyo Institute of Technology, 4259 Nagatsuta, Yokohama 226-8501, Japan}
\affiliation{Department of Materials Science and Engineering, Tokyo Institute of Technology, 4259 Nagatsuta, Yokohama 226-8501, Japan}

\author{Yosuke Kayanuma}
\affiliation{Laboratory for Materials and Structures, Institute of Innovative Research, Tokyo Institute of Technology, 4259 Nagatsuta, Yokohama 226-8501, Japan}
\affiliation{Graduate School of Sciences, Osaka Metropolitan University, 1-1 Gakuen-cho, Sakai, Osaka, 599-8531 Japan}

\author{Kazutaka G. Nakamura}
\email[Corresponding author: ]{nakamura@msl.titech.ac.jp}
\affiliation{Laboratory for Materials and Structures, Institute of Innovative Research, Tokyo Institute of Technology, 4259 Nagatsuta, Yokohama 226-8501, Japan}
\affiliation{Department of Materials Science and Engineering, Tokyo Institute of Technology, 4259 Nagatsuta, Yokohama 226-8501, Japan}

\date{\today}

\begin{abstract}
    We present a novel and simple picture of the generation dynamics of coherent longitudinal optical (LO) phonons and LO-phonon-plasmon-coupled (LOPC) modes by the ultrafast infrared pump-pulses in gallium arsenide (GaAs) employing the low-temperature approximation.
    LO phonons exhibit a pronounced coupling with plasmons formed by the optically excited electrons in the excited states of GaAs. 
    This coupling results in the coherent oscillation of the LOPC modes in the excited states. 
    The pump pulse also induces stimulated Raman scattering, which generates the coherent LO-phonon oscillation in the ground state. 
    This picture is incorporated into a simplified model, and the time evolution of the density operator is calculated using the Lindblad-type quantum master equation. 
    The theoretical results explain well the reported experimental results on the coherent oscillation of LO phonons and LOPC modes observed through transient reflection measurements. 
    Above all, our model provides a natural reason for the simultaneous manifestation of the LO phonons and the LOPC modes.
\end{abstract}

\maketitle

\section{INTRODUCTION}
The interplay of quasiparticles in solids plays an essential role in a wide range of physical phenomena \cite{Mattuck1992, Mahan2000}.
For instance, electron-phonon interactions hold an important position in condensed matter physics \cite{Ziman1960, Cardona2005} and are also critical in discussions concerning ultrafast phenomena.
In certain metallic and semiconducting materials, the plasmon appears as a quasiparticle associated with the collective motion of electrons \cite{Bohm1953, Dpines2018}.
In polar semiconductors such as GaAs, LO-phonon-plasmon-coupled (LOPC) modes form \cite{Varga1965, Singwi1966, Peschke1993, Kuznestov1995}, and have been observed by both Raman spectroscopy \cite{Mooradian1966, Abstreiter1979} and coherent phonon spectroscopy with transient reflectivity measurements \cite{Cho1996, Hase1999, Ishioka2011, Hu2012}. 
The frequency of the plasmon is proportional to the square root of the conduction electron density.
If the plasmon frequency is close to that of the LO phonon, new modes denoted $L_+$ and $L_-$ are generated via the LO-phonon-plasmon interaction, the eigenstates of which are obtained through the diagonalization of the interaction Hamiltonian \cite{Singwi1966, Peschke1993}.

Nevertheless, it is often reported that the bare LO phonon oscillation continues to manifest in transient reflection measurements along with the $L_+$ and $L_-$ modes \cite{Hase1999, Ishioka2011, Hu2012} and in Raman spectroscopy \cite{Abstreiter1979}.
This is strange because the LOPC modes are real eigenmodes in the excited states and the LO mode is simply its building block.
The simultaneous coexistence of the LO phonon and LOPC modes has been discussed in connection with an inhomogeneous density distribution of electrons --for example, a surface depletion layer for a depth distribution of doped electrons \cite{Abstreiter1979, Ishioka2011} and an in-plane distribution of excitation densities produced by focusing the laser beam tightly on the sample \cite{Kuznestov1995, Cho1996}.
In this model, the LOPC modes and the bare LO phonons are assumed to be spatially segregated depending on the electron density in the crystal.

Coherent optical phonons are generated by irradiating ultrashort optical pulses shorter than the phonon period \cite{Forst2008, Shinohara2010}.
The coherent phonon oscillations are generated not only during the absorption process but also in the stimulated Raman process even under opaque conditions \cite{Nakamura2015}.
The present authors have performed a study of the generation processes of coherent LO phonons in GaAs in pump--probe measurements employing the double pulse pumping technique using a pair of optical pulses with a relative phase controlled to within an attosecond accuracy \cite{Nakamura2019, Takagi2023-2}.
This technique enabled us to clarify the ultrafast generation processes of the coherent phonon through quantum-path interference.
An important result is that, with no contribution from the absorption process, the coherent LO phonons are generated through the Raman process.

In the present work, we propose a different and simple model to explain the spectral coexistence of the LO mode and the LOPC modes in coherent phonon generation within a GaAs crystal.
The amplitude of the coherent phonons is calculated by solving the Lindblad-type quantum master equation while adopting a low-temperature approximation.
The almost bare LO phonons are shown to be generated in the ground state through a stimulated Raman process, whereas the LOPC modes are generated in the excited states through the impulsive absorption process.

\section{THEORETICAL MODEL}
We consider a composite system of electrons, optical phonons, and a plasmon system, and focus on coherent phenomena.
In this model, and within the low-temperature approximation, we assume that coherently photo-excited electrons interact with phonons and plasmons in which the doped electron density is negligibly small.

The wavelength of the optical pulse is much longer than the lattice constant, and hence the electron--hole pairs are uniformly distributed near the $\Gamma$-point.
Therefore, optical phonons need only be considered near the $\Gamma$-point ($q \approx 0$).
We modify the standard Hamiltonian for an electron--phonon--plasmon system \cite{Singwi1966,Peschke1993, Lee2008, Lee2009,Sakata2021}.
The Hamiltonian used in this paper is described as follows (details are given in \textcolor{black}{Supplemental Material A});
\begin{eqnarray}
    H_{0} &=& \varepsilon_{g}\ket{g}\bra{g} + \varepsilon_{e}\ket{e}\bra{e} + \hbar \omega_{\rm{ph}}b^{\dagger}b    \nonumber\\
    &+& \alpha \hbar \omega_{\rm{ph}}(b+ b^{\dagger})\ket{e}\bra{e} \nonumber\\
    &+& \sum_{\eta=g,e} \hbar \omega_{\rm{pl,\eta}} c_\eta^{\dagger}c_\eta \ket\eta  \bra{\eta}  \nonumber\\
    &+&  \gamma \sum_{\eta=g,e}  \hbar \sqrt{\omega_{\rm{ph}}  \omega_{\rm{pl},\eta}}(b+ b^{\dagger})(c_\eta+ c_\eta^{\dagger})\ket{\eta}\bra{\eta}.
\end{eqnarray}
The parameters $\omega_{\rm{ph}}$, $\omega_{\rm{pl}}$ represent the LO phonon and plasmon frequencies, respectively.
For n-doped semiconductors, the plasmon frequency in the conduction band depends on the donor induced and photo-excited electrons ($n$).
The density of the thermally excited electrons from the dopant is approximately $10^{17} {\rm{cm^{-3}}}$ for the Si-doping density of $10^{18} {\rm{cm^{-3}}}$ at 90 K, and its plasmon--phonon coupling maybe negligible.
Moreover, in the valence band, the plasmon frequency is dependent on the donor induced electrons, which is negligible at low temperatures.
We also set $\omega_{\rm{pl, e}} = \sqrt{n e^{2}/m^{*}\epsilon_{\infty} \epsilon_0}$ and $\omega_{\rm{pl, g}} = 0$, where $n$ is the electron density, $m^*$ the effective mass of an electron, $e$ the elementary charge, and $\epsilon_0$ the vacuum permittivity.
The parameter $\alpha$ represents the dimensionless coupling constant of the electron--phonon system.
In the bulk, the Huang--Rhys factor ($\alpha^{2}$) is considered very small ($\alpha^{2}\ll1$).
The parameter $\gamma$ denotes the coupling constant between phonon and plasmon and is defined by $\gamma=\frac{1}{2}\sqrt{1-\epsilon_{\infty}/\epsilon_{s}}$, with high-frequency and static-dielectric constants: $\epsilon_{\infty}$ and $\epsilon_{s}$.
In the limit of large electron density $n$, the frequency of $L_-$ is known to approaches the frequency of the transversal optical phonon ($\omega_{TO}$) \cite{Varga1965,Singwi1966,Peschke1993}. 
The operators $b^{\dagger}$ and $b$ denote, respectively, the creation and annihilation operators of an  optical phonon at the $\Gamma$-points.
Similarly, the operators $c^{\dagger}$ and $c$ denote the creation and annihilation operators of the plasmon.

Adopting the \textcolor{black}{rotating--wave} approximation (RWA) and the dipole interaction, the light--electron interaction Hamiltonian $H_{I}(t)$ is expressed as
\begin{equation}
    H_{I}(t) = \mu E(t) e^{-i\Omega_{0}t} \ket{e}\bra{g} + H.c.,
\end{equation}
where $\mu$ denotes the transition dipole moment, $\Omega_{0}$ the frequency of the optical pulse, and
$E(t)$ the optical envelope of the pump pulse, represented by $E(t)=E_{0}f(t)$ for a single pulse.
The function $f(t)$ represents a Gaussian envelope function with a width of $\sigma$, and $E_{0}$ the amplitude of the electric field.

We consider the system of interest (the electron--phonon--plasmon composite system) to interact with the environment including the incoherent phonons, electrons, and impurities, and treat it as an open quantum system \cite{Nakamura2024}.
The time-evolution of the quantum state is derived from the Lindblad-type master equation \cite{Breuer2007}.
\textcolor{black}{
\begin{eqnarray}
    i \hbar \frac{d}{dt}\rho (t) &=& [H, \rho(t)] +i\hbar \hat{D}(\rho(t)),\\
    \hat{D}(\rho(t)) &=& \sum_{i} \Gamma_{i} \left[ L_{i} \rho(t) L_{i}^{\dagger} - \frac{1}{2}\{L_{i}^{\dagger}L_{i},\rho(t)\}\right],
\end{eqnarray}
}
where $H$ denotes the total Hamiltonian ($H = H_{0}+H_{I}(t)$), and $\rho(t)$ the density operator $\rho(t) = \ket{\psi(t)}\bra{\psi(t)}$.
Moreover, $\hat{D}(\rho(t))$ and $L_{i}$ denote the dissipator and the Lindblad operator, and $\Gamma_{i}$ denotes the the decay rate.
We define the phase relaxation of the electron, phonon, and plasmon states as the Lindblad operator $L_{i}$ in Eq.~(4); specifically, \ $L_{{\rm{el}}} = \ket{e}\bra{e}, $\ $L_{{\rm{ph}}} = b^{\dagger}b$,  and $L_{{\rm{pl}}} = c^{\dagger}c$.

The wave function $\ket{\psi(t)}$ of the composite systsm is expanded using the basis sets of the electron, phonon, and plasmon subsystems,
\begin{eqnarray}
   \ket{\psi(t)} 
   &=& \sum_{n,m} \left(c_{n,m}(t)\ket{g,n,m} + d_{n,m}(t)\ket{e,n,m}\right),
\end{eqnarray}
where $\ket{n}$, and $\ket{m}$ denote the phonon and plasmon Fock states, respectively; 
$c_{n,m}(t)$ and $d_{n,m}(t)$ the time-dependent probability amplitudes of the $n$-phonon and $m$-plasmons correlated with the electronic ground and excited states, respectively.
The initial state is set to an electronic ground state coupled with the states of zero phonons and zero plasmons, $\ket{\psi(0)} = \ket{g,0,0}$, \textcolor{black}{which is a separable state}.

We calculate the expectation value of the phonon nuclear coordinates using the time evolution of the density operator obtained from the master equation ($\braket{Q(t)}={\rm{Tr}}\left[Q\rho(t)\right]$, where \ $Q = \sqrt{\hbar/2M\omega_{\rm{LO}}}(b^{\dagger}+b)$ with $M$ the reduced mass of atoms per unit cell of GaAs).
The coherent phonon behavior is governed by the coherence term of the density operator.
In our model, we can evaluate the expectation value of the phonon coordinates by selecting the associated electronic state ($\ket{g}$\ or $\ket{e}$).
Therefore, we evaluate the expectation value under three conditions;
$\braket{Q(t)} = \braket{Q_{\rm{g}}(t)} +\braket{Q_{\rm{e}}(t)}$,
\ $\braket{Q_{\rm{g}}(t)} = \sum_{i,j,m} \braket{g,i,m|Q|g,j,m}$,\ and 
$\braket{Q_{\rm{e}}(t)} = \sum_{i,j,m} \braket{e,i,m|Q|e,j,m}$.

\begin{figure}[]
   \centering
   \includegraphics[scale=0.68]{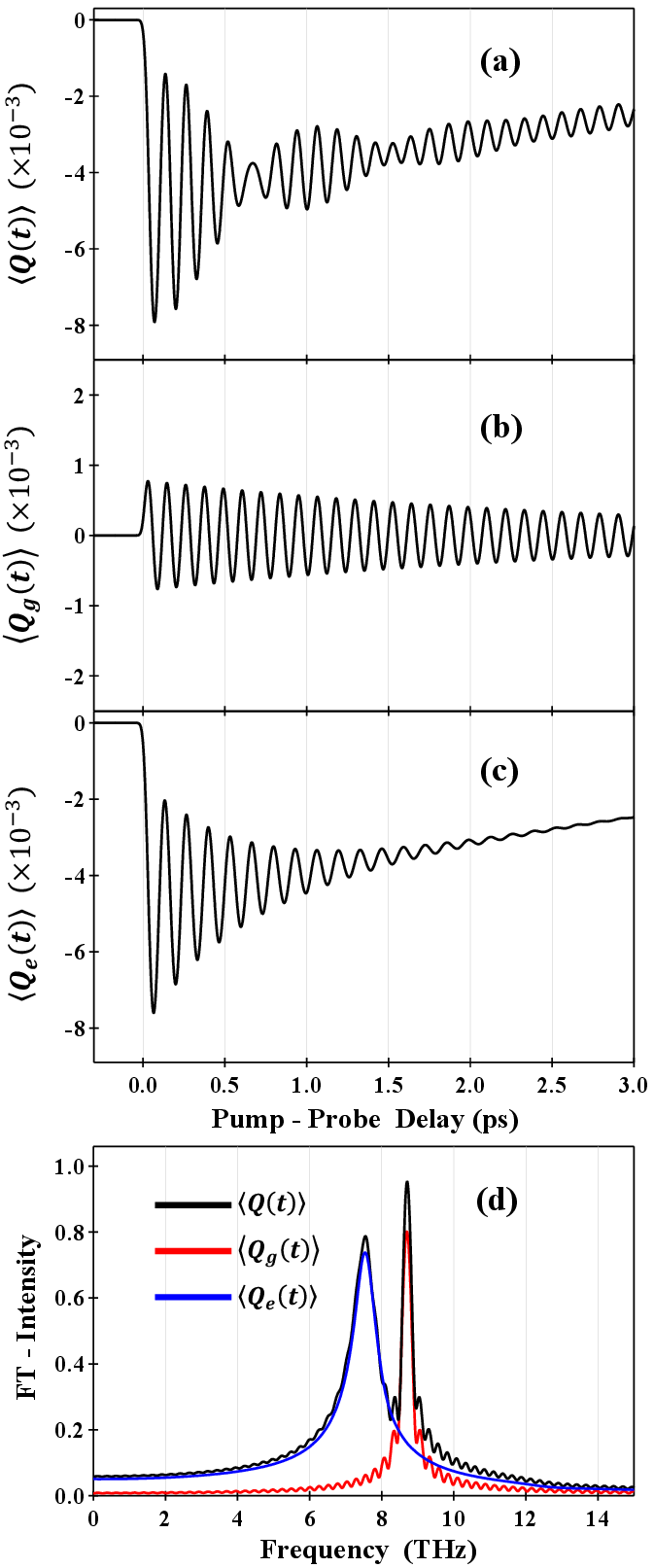}
   \caption{Theoretical calculation of displacement $\braket{Q(t)}$ of the LO phonons: (a) for the sum of electronic ground and excited states, (b) for the electronic ground state, and (c) for the electronic excited states as a function of the pump--probe delay.
   (d) Fourier spectra of each phonon expectation value obtained from the calculation; the black, blue, and red lines correspond, respectively, to the sum of the electronic ground and excited states, the electronic ground state, and the electronic excited state.
   The range of the Fourier transform is set with a delay of 0.1 -- 4.0 ps.
   The non-vibrating component has not been removed.}
   \label{Fig1}
\end{figure}

\begin{figure}[]
    \centering
    \includegraphics[scale=0.38]{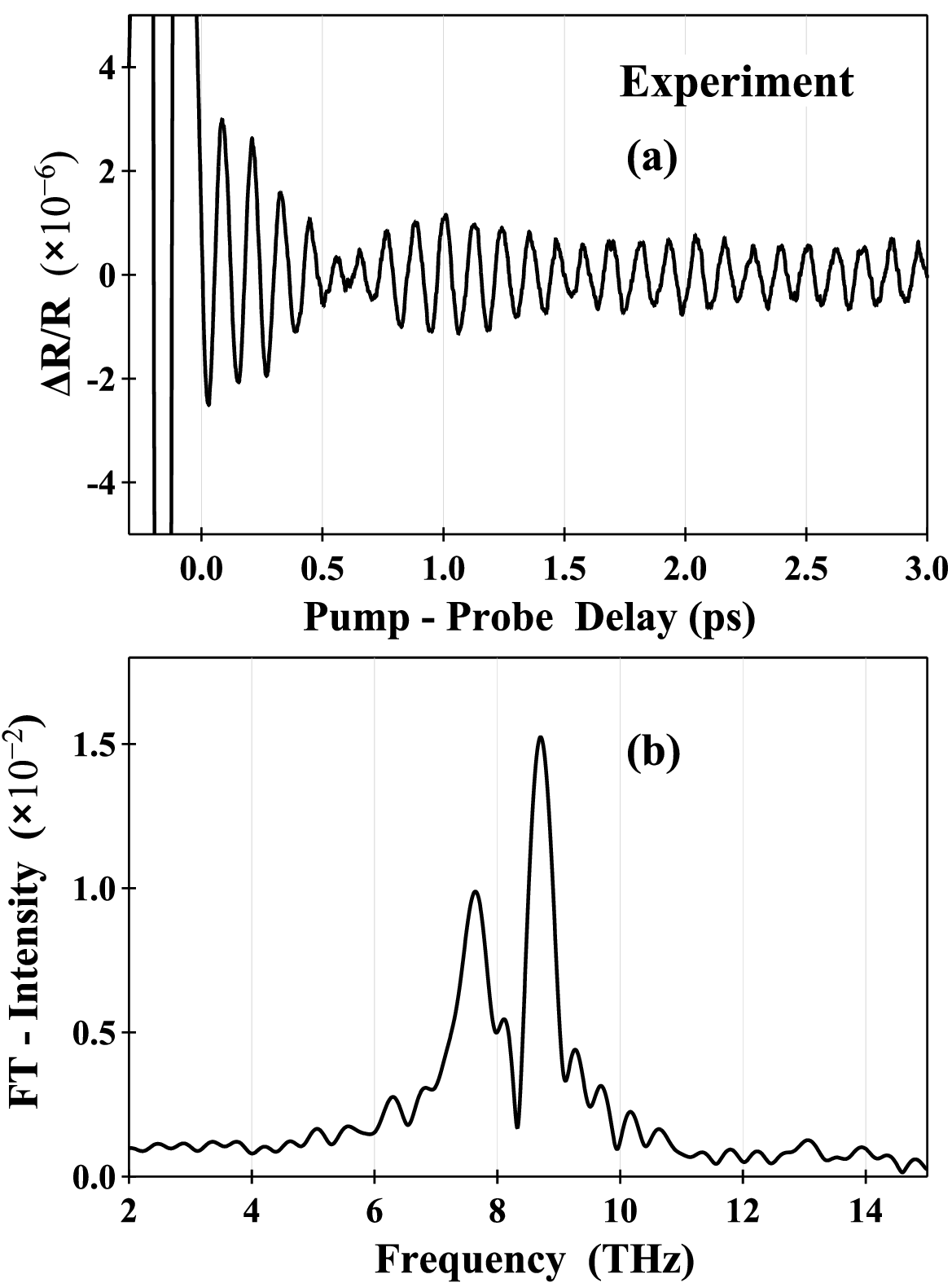}
    \caption{Experimental coherent phonon measurements (transient reflectance, $\Delta R/R$) for n-GaAs (with a Si-doping density of $1.0 \times 10^{18}\ {\rm{cm^{-3}}}$).
    (a) Time evolution of transient reflectance ($\Delta R/R$) taken at a temperature of 90 K. 
    Coherent artifacts and non-vibrating background signals were removed.
    (b) Fourier spectra of transient reflectance ($\Delta R/R$) in pump-probe delay times ranging from 0.15 to 2.50 ps.}
 \end{figure}

In the actual calculation, we used the Lindblad-type master equation implemented in the QuTiP solver of the Python package \cite{Qutip}.
We used two electronic states and three Fock states for both the phonon and plasmon.
The electron--phonon coupling constant $\alpha$ is set to 0.01.
The bandgap energy is set to $\varepsilon_{e}-\varepsilon_{g}$ = 1.52 eV which corresponds to the center of the optical frequency $\Omega_{0}$. 
The vibrational energy of the LO-phonon is set to $\hbar \omega_{\rm{ph}}$ = 36 meV (approximately 8.7 THz). 
The high- and static-frequency dielectric constants are set to $\epsilon_{\infty}=11.1$ and $\epsilon_{s}=13.1$, respectively \cite{Cardona2005}.
The optical pulse parameter is set to $\sigma$ = 30 fs.
The transition strength ($\mu E_{0}$) is determined from settings of the laser pulse used in the previous work \cite{Takagi2023}.
\textcolor{black}{Furthermore, we set the decay rate of each phase-relaxation as $\Gamma_{{\rm{el}}}=2/23\ (1/{\rm{fs}})$, $\Gamma_{{\rm{ph}}}=2/3000\ (1/{\rm{fs}})$, and $\Gamma_{{\rm{pl}}}=2/150\ (1/{\rm{fs}})$.
In this case, the coherent terms, calculated with Eqs. (3) and (4),  decrease exponentially with $\Gamma_i /2$.}
The decay constant $\Gamma_{{\rm{ph}}}$ is set based on the transient reflectance measurement of n-GaAs \cite{Ishioka2011, Hu2018}.
Additionally, the decay constant $\Gamma_{{\rm{el}}}$ is set based on the measured electronic decoherence time \cite{Takagi2023-2, comment1}. \\

The amplitude of the electric field is estimated as:
\begin{eqnarray}
    E_{0} = \sqrt{\frac{\sigma}{c \epsilon_0} F \sqrt{\frac{\pi}{2}} },
\end{eqnarray}
where $F$ denotes the fluence [J/cm$^{2}$] of the pump pulse, $c$ the speed of light, and $\epsilon_0$ the dielectric constant of vacuum.
The photo-induced electron density is calculated from $n_e = \beta F$ given the absorption coefficient $\beta$ [cm$^{-1}$].
For simplicity, the reflectivity of GaAs is not considered in this calculation.
The transition dipole moment $\mu$ is estimated using $\mu = q \times r$, where $q$ is the charge and $r$ is the distance between Ga and As atoms. 
For the calculation, we use an effective value of $\mu = q \times r/2$.


\section{RESULTS AND DISCUSSION}
Initially, we evaluated the time evolution of coherent phonon oscillation under single-pulse excitation (25.0 $\mu {\rm J/ cm}^2$).
Figure \ref{Fig1} illustrates the three conditions of the expectation value for the phonon nuclear coordinates ($\braket{Q(t)}$,\ $\braket{Q_{g}(t)}$, and $\braket{Q_{e}(t)}$), in which the non-vibrating component has not been removed.
From Fig.~1(a), the total coherent oscillation $\braket{Q(t)}$ exhibits beating, consisting of two oscillations with frequencies 7.5 and 8.7 THz, which are attributed to the lower branch of the LO-phonon-plasmon coupling ($L_{-}$) and LO phonons, respectively.
In contrast, $\braket{Q_g(t)}$ shows a single damped oscillation at a frequency of 8.7 THz (Fig.~1 (b)).
The relative phase of these oscillations for the LO and LOPC-modes is $\pi/2$;
it is clearly characteristic of Raman and absorption processes \cite{Nakamura2015}.
Moreover, in Fig.~1(c), the coherent oscillations at electronic excited state ($\braket{Q_{e}(t)}$) also show a single damped oscillation with a frequency of 7.5 THz.
The damping behavior is faster than that of the electronic ground state illustrated in Fig.~1(b).
\textcolor{black}{The different coherent oscillations indicate, depending on the electronic state, that the electronic and phonon states are entangled.}

We performed Fourier transformation of Fig.~1(a)--(c), with pump-probe delay ranging from 0.1 to 4.0 ps, respectively (Figure 1(d)).
The Fourier spectra indicate that a LO-phonon oscillation is found only when we select the electric ground state.
The $L_{-}$ mode oscillation is found when the LO phonons are coupled to the electronic excited state.
LO phonons play a dual role as bare phonons in the ground state and as plasmon-coupled mode in the excited state.
The peak of the higher frequency $L_{+}$ mode (approximately 11 THz) is too small to be seen in  Fig.~1(d).
We found that the dip in the total phonon spectrum at 8.5 THz is due to the phase difference between LO-phonon and LOPC ($L_{-}$) mode \cite{Nakamura2015}.
This dip has also been reported by experiment \cite{Ishioka2011}.

Figure 2(a) presents the experimental transient reflectance measurements of n-GaAs (with a Si-doping density of $1.0 \times 10^{18}\ {\rm{cm^{-3}}}$) taken at 90K using a Ti: sapphire laser with a center frequency of 800 nm.
The coherent artifacts and non-vibrating background signals were removed.
Additionally, the time origin of the pump--probe delay (horizontal axis) was set to 0 fs at the beginning of the oscillation after the coherent artifact.
The oscillation behaviors in Figs.~2(a) and Fig.~1(a) are very similar.
The Fourier spectrum obtained from the experiment [Fig. ~2(b)] shows $L_{-}$ and LO-phonon signals at approximately 7.6 and 8.7 THz, respectively, but no $L_{+}$ signal was observed between 11 - 12 THz.
The $L_{+}$ signal is not observed in the weak excitation intensity region of the pump pulse.

\begin{figure}[]
   \centering
   \includegraphics[scale=0.4]{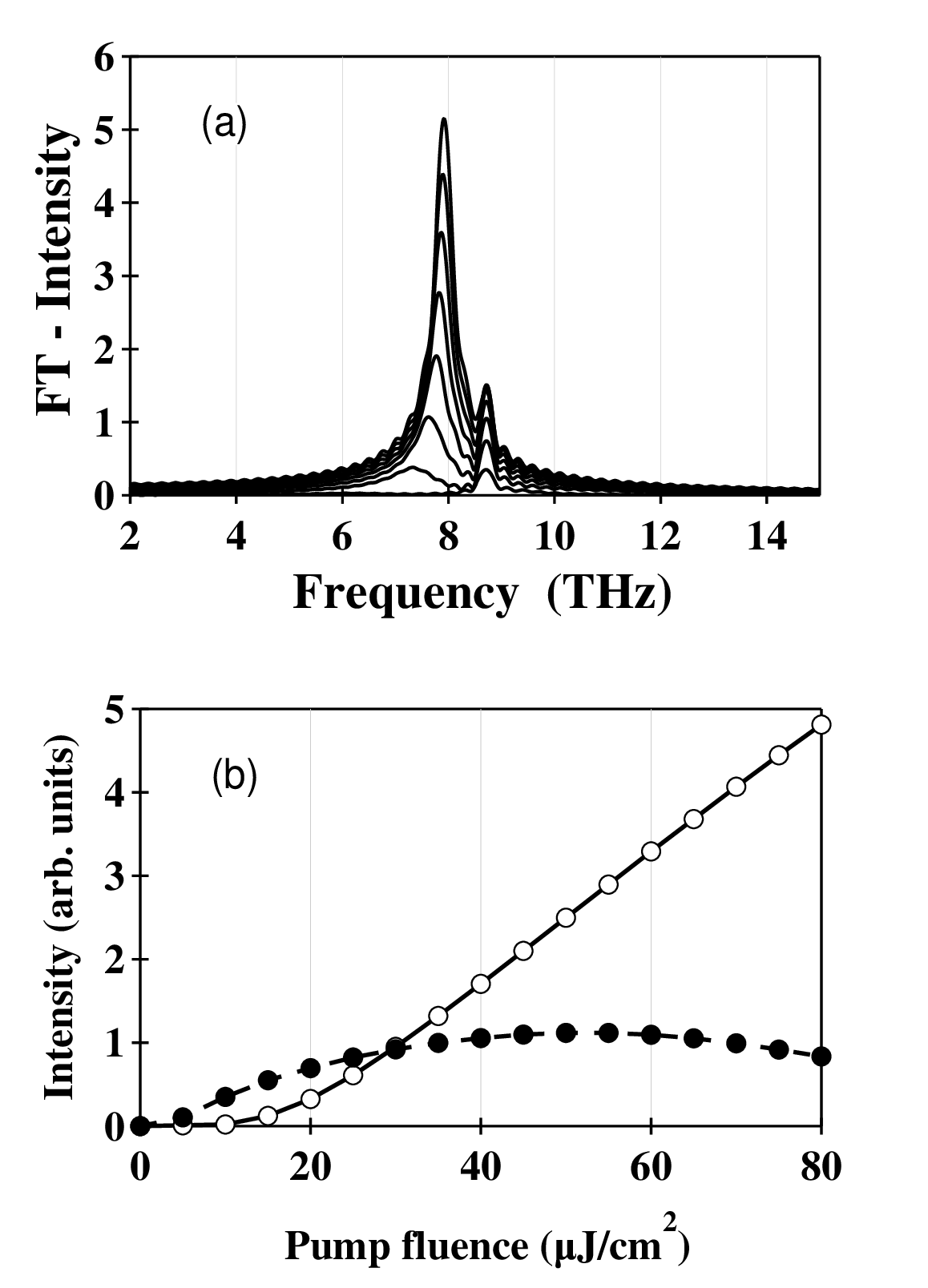}
   \caption{Comparison of the laser fluence of the pump pulse with the theoretical Fourier spectrum.
   We calculated the fluence between 10 and 80~$\mu$J/cm$^{2}$ in steps of \textcolor{black}{5} $\mu$J/cm$^{2}$.
   {(a) Fourier spectrum as a function of frequency \textcolor{black}{(10 and 80~$\mu$J/cm$^{2}$ in steps of 10 $\mu$J/cm$^{2}$)}; (b) intensities of LO phonons (black circles) and L$_-$ (white circles) as a function of pump fluence.}}
\end{figure}

We also evaluated from theory the excitation intensity dependence on the generation efficiency for the LO phonon and LOPC modes [Fig.~3(a)] up to a fluence of 80 $\mu$J/{cm$^{2}$}.
The oscillation intensity of the LO phonon shows nonlinear behavior as laser fluence increases (Fig. 3 (b)) 
In contrast, the LOPC-mode oscillation intensity changes almost linearly up to about 80 $\mu$J/{cm$^{2}$}.
These features have also been observed in experiments \cite{Ishioka2011}.
The differences in phonon generation behavior can be explained with results from previous work \cite{Takagi2023} as being due to differences in quantum paths, namely, coherent phonon generation via impulsive stimulated Raman and impulsive absorption processes.
The results show that the coherent LO phonons behave as a bare LO phonon oscillation in the ground state and the LOPC coupled oscillation in the excited state.

To summarize, we proposed that the coexistence of LO and LOPC modes observed in the optical pumping experiments of GaAs crystal is attributed to the difference in the quantum paths---the LO mode by the impulsive stimulated Raman path and the LOPC mode by the impulsive absorption path.
We calculated the probabilities associated with the LO phonons and LOPC modes in n-type GaAs by solving the Lindblad-type master equation for the simple model Hamiltonian describing the electron--phonon--plasmon system.
We applied the low-temperature approximation in which the density of the thermally excited electrons from dopants is negligibly small and the phonon--plasmon coupling in the electronic ground state is negligible.
The interpretation is consistent with the experimental data obtained.
Our model provides a natural explanation for the simultaneous manifestation of the LO phonons and the LOPC modes.

\begin{acknowledgments}
We thank  R.~Takai and M.~Hirose for their help for experiment.
KGN thanks Gordon Han Ying Li for his discussion on model for LOPC at the early stage of this work.
This work was partially supported by JSPS KAKENHI under Grant numbers 19K03696, 19K22141, 21K18904, 22J23231, 22H01984, 22KJ1342; and \textcolor{black}{23K23252} and by Design \& Engineering by Joint Inverse Innovation for Materials Architecture; MEXT, and 
Collaborative Research Projects of Laboratory for Materials and Structures, Institute of Innovative Research, Tokyo Institute of Technology. 
\end{acknowledgments}



\begin{thebibliography}{99}
\bibitem{comment1} The decay rates are dependent on an electron density. However, in the present calculation, we use constant decay rates.
\end{thebibliography}
\end{document}


\title{\textcolor{black}{Supplemental Material}: Dual role of longitudinal optical phonons for generation of coherent oscillations in gallium arsenide under optical pumping}

    \author{Itsuki Takagi}
    \affiliation{Laboratory for Materials and Structures, Institute of Innovative Research, Tokyo Institute of Technology, 4259 Nagatsuta, Yokohama 226-8501, Japan}
    \affiliation{Department of Materials Science and Engineering, Tokyo Institute of Technology, 4259 Nagatsuta, Yokohama 226-8501, Japan}
    
    \author{Yuma Konno}
    \affiliation{Laboratory for Materials and Structures, Institute of Innovative Research, Tokyo Institute of Technology, 4259 Nagatsuta, Yokohama 226-8501, Japan}
    \affiliation{Department of Materials Science and Engineering, Tokyo Institute of Technology, 4259 Nagatsuta, Yokohama 226-8501, Japan}
    
    \author{Yosuke Kayanuma}
    \affiliation{Laboratory for Materials and Structures, Institute of Innovative Research, Tokyo Institute of Technology, 4259 Nagatsuta, Yokohama 226-8501, Japan}
    \affiliation{Graduate School of Sciences, Osaka Metropolitan University, 1-1 Gakuen-cho, Sakai, Osaka, 599-8531 Japan}
    
    \author{Kazutaka G. Nakamura}
    \email[Corresponding author: ]{nakamura@msl.titech.ac.jp}
    \affiliation{Laboratory for Materials and Structures, Institute of Innovative Research, Tokyo Institute of Technology, 4259 Nagatsuta, Yokohama 226-8501, Japan}
    \affiliation{Department of Materials Science and Engineering, Tokyo Institute of Technology, 4259 Nagatsuta, Yokohama 226-8501, Japan}
    
    \date{\today}
    
\maketitle

\section{A.\ Details on the treatment of theoretical models}
Here, we describe the details of the quantum model for the simplified electron-phonon-plasmon composite system that is treated in this paper.
The system Hamiltonian for the electron--phonon--plasmon composite system is described as follows;
\begin{eqnarray}
        H_{0} &=& H_{\rm{el}} + H_{\rm{ph}} + H_{\rm{pl}} + H_{\rm{el-ph}} + H_{\rm{ph-pl}} + H_{\rm{el-pl}}\\
        H_{\rm{el}} &=& \sum_{k} \sum_{v = h,l} \varepsilon_{v}(k)a_{k,v}^{\dagger} a_{k,v} + \sum_{k}\varepsilon_{c}(k)a_{k,c}^{\dagger}a_{k,c},\\
        H_{\rm{ph}} &=& \hbar \omega_{\rm{ph}} b^{\dagger}b,\\
        H_{\rm{pl}} &=& \sum_{k} \sum_{v = h,l}\hbar \omega_{\rm{pl,v}} c_{v}^{\dagger}c_{v} \ a_{k,v}^{\dagger} a_{k,v} + \sum_{k} \hbar \omega_{\rm{pl,c}} c_{c}^{\dagger}c_{c} \ a_{k,c}^{\dagger} a_{k,c},\\
        H_{\rm{el-ph}} &=& \sum_{k} \alpha \hbar \omega_{\rm{ph}} (b + b^{\dagger}) a_{k,c}^{\dagger}a_{k,c}, \\
        H_{\rm{ph-pl}} &=& \sum_{k} \gamma \sum_{\eta = v, c}  \hbar \sqrt{\omega_{\rm{ph}}\omega_{\rm{pl,{\eta}}}}(b+b^{\dagger})(c_{\eta}+c_{\eta}^{\dagger}) a_{k,\eta}^{\dagger}a_{k,\eta},\\
        H_{\rm{el-pl}} &=& \sum_{k} \beta \hbar \omega_{\rm{pl, {c}}} (c_{c}^{\dagger} + c_{c}) a_{k,c}^{\dagger}a_{k,c}, 
\end{eqnarray}
where $H_{\rm{el}}$, $H_{\rm{ph}}$, and $H_{\rm{pl}}$ respect the Hamiltonian of the electron, LO phonon, and plasmon system, respectively.
Also, $H_{\rm{el-ph}}$ and $H_{\rm{ph-pl}}$ denote the electron--phonon interaction and the phonon--plasmon interaction Hamiltonian, respectively.

The parameter $\omega_{\rm{ph}}$ is the LO phonon frequency. 
The parameters $\omega_{\rm{pl,{v}}}$ and $\omega_{\rm{pl,{c}}}$ are the plasmon frequencies in valence and conduction bands, and are defined by $\omega_{\rm{pl}} = \sqrt{n e^{2}/m^{*}\epsilon_{\infty} \epsilon_0}$ with high-frequency dielectric constant $\epsilon_{\infty}$, effective electron mass $m^{*}$, elementary charge $e$, $\epsilon_0$ the vacuum permittivity, and electron density $n$.
In the case of an n-doped semiconductor, the plasmon frequency in the conduction band ($\omega_{\rm{pl,{c}}}$) depends on the conduction electron density (with e.g., donor-induced and photo-induced electrons). 
Also, the plasmon frequency in the valence band ($\omega_{\rm{pl,{v}}}$) is determined by only the dopant-induced electron density.
\textcolor{black}{In the Raman process, photo-induced electrons are once excited to the conduction band as an intermediate state, but they instantly return to the valence band. Therefore, the photo-induced electron density does not affect the plasmon in the valence band (i.e., the electronic ground state as we define it).}

The parameter $\alpha$ is the dimensionless coupling constant of the electron-phonon system.
In the bulk case, the Huang-Rhys factor ($\alpha^{2}$) is considered very small ($\alpha^{2}\ll1$).
\textcolor{black}{Also, the $\beta$ is the dimensionless coupling constant of the electron-plasmon system.
The electron--plasmon interaction is very small compared to the electron-phonon interaction \cite{Peschke1993}.
Therefore, in this main paper, we do not consider the influence of electron--plasmon interaction, and treat as $\beta \approx 0$.}
\textcolor{black}{$\gamma$ is the dynamical coupling constant between phonon and plasmon under conduction or valence band, and is defined by $\gamma=\frac{1}{2}\left(1-\epsilon_{\infty}/\epsilon_{s}\right)^{1/2}$, with static dielectric constant $\epsilon_{s}$.}
\textcolor{black}{In the low-temperature approximation, the phonon--plasmon interaction in valence bands is not affected effectively, and treated as $\gamma \hbar \sqrt{\omega_{\rm{ph}}\omega_{\rm{pl,{v}}}}(b+b^{\dagger})(c_{v}+c_{v}^{\dagger}) a_{k,v}^{\dagger}a_{k,v} \approx 0$.}

The operators $a_{k,v}$ and $a_{k,c}$ are the annihilation operators for electrons in the valence and conduction bands, and $a^{\dagger}_{k,v}$, $a^{\dagger}_{k,c}$ are the creation operator. 
In addition, the subscript $v$ represents the heavy hole ($h$) and light hole ($l$) bands.
The operators $b^{\dagger}$, $b$ are the creation and annihilation operators of optical phonon at $\Gamma$-points, respectively.
Also, the operators $c^{\dagger}$, $c$ are the creation and annihilation operators of the plasmon, respectively.

The light--electron interaction Hamiltonian $H_{I}(t)$ can be described by using the \textcolor{black}{rotating} wave approximation (RWA) and the dipole interaction;
\begin{equation}
    H_{I}(t) = \sum_{k}\sum_{v = h,l}\mu_{k,v} E(t) e^{-i\Omega_{0}t} a_{k,c}^{\dagger}a_{k,v} + H.c.,
\end{equation}
where the parameters $\mu_{k, v}$ is the transition dipole moment and $\Omega_{0}$ is the frequency of the optical pulse.
$E(t)$ is the optical envelope of the pump pulse, represented by $E(t)=E_{0}f(t)$ for a single pulse.
The function $f(t)$ is a Gaussian envelope function with a width of $\sigma$, and $E_{0}$ is an amplitude of electric field.

Here, the electronic state in the valence band is treated as a simplified state (i.e., the electronic ground state $\ket{g}$). 
As a treatment, it is assumed that the initial electron distribution occupies the valence band, e.g., by the low-temperature approximation.
Therefore, we define the electronic state in valence band as $\ket{g} =\Pi_{k,v} a^{\dagger}_{k,v}\ket{vac}$, where $\ket{vac}$ is the vacuum of the electron.
We also define the electronic (excited) state in conduction band as $\ket{e} = \sum_{k}\sum_{v=h,l} \ket{e_{k,v}} = \sum_{k}\sum_{v=h,l} a_{k,c}^{\dagger}a_{k,v}\ket{g}$.
In this model, we consider only representative excited states ($\hbar \Omega_{0} = \varepsilon_{c}(k)-\varepsilon_{v}(k)$) with a wavevector whose bandgap matches the optical excitation energy in the perspective of RWA.

From the above, we simplify the Hamiltonian in Eqs.~(1)-(6) as follows;
\begin{eqnarray}
    H_{0} &=& H_{\rm{el}} + H_{\rm{ph}} + H_{\rm{pl}} + H_{\rm{el-ph}} + H_{\rm{ph-pl}}\\
    H_{\rm{el}} &=& \varepsilon_{g}\ket{g}\bra{g} + \varepsilon_{e}\ket{e}\bra{e},\\
    H_{\rm{ph}} &=& \hbar \omega_{\rm{ph}} b^{\dagger}b,\\
    H_{\rm{pl}} &=& \sum_{\eta = g, e} \hbar \omega_{\rm{pl,{\eta}}} c_{\eta}^{\dagger}c_{\eta} \ket{\eta}\bra{\eta}, \\
    H_{\rm{el-ph}} &=&  \alpha \hbar \omega_{\rm{ph}} (b + b^{\dagger}) \ket{e}\bra{e}, \\
    H_{\rm{ph-pl}} &=& \gamma \sum_{\eta = g, e}  \hbar \sqrt{\omega_{\rm{ph}}\omega_{\rm{pl,{\eta}}}}(b+b^{\dagger})(c_{\eta}+c_{\eta}^{\dagger}) \ket{\eta}\bra{\eta},
\end{eqnarray}